\documentclass[aps,pra,showpacs,amsmath,amssymb,twocolumn]{revtex4}

\newcommand{\tfrad}[1]{R_{#1}}
\newcommand{\pf}{p_f}

\usepackage{epsfig}

\begin{document}
\title{Trapped Phase-Segregated Bose-Fermi Mixtures and Their Collective
 Excitations}
\author{Bert Van Schaeybroeck and Achilleas Lazarides} \affiliation{Instituut
 voor Theoretische Fysica,\\ Katholieke Universiteit
Leuven,
 Celestijnenlaan 200 D, B-3001 Leuven, Belgium.}


\begin{abstract}
Recent progress in the field of ultracold gases has allowed the
creation of phase-segregated Bose-Fermi systems. We present a
theoretical study of their collective excitations at zero
temperature. As the fraction of fermion to boson particle number
increases, the collective mode frequencies take values between
those for a fully bosonic and those for a fully fermionic cloud,
with damping in the intermediate region. This damping is caused by
fermions which are resonantly driven at the interface.
\end{abstract}

\maketitle
\section{Introduction}
Since their first experimental observation~\cite{truscott},
degenerate boson-fermion (BF) systems have been realized using
more than five different particle mixtures~\cite{schreck}. The one
composed of fermionic $^{40}$K and bosonic $^{87}$Rb particles is
the most thoroughly studied: experiments have been performed on
its collective excitations~\cite{ferlaino2} and on collapse and
expansion dynamics~\cite{ferlaino3,ospelkaus2}. Using a Feshbach
resonance to tune the interspecies interaction towards repulsion
produced strong signs of phase
segregation~\cite{zaccanti,ospelkaus}; the presence of such
resonances has also been observed in BF mixtures of different
species~\cite{deh}. Despite these advances, many new physical
phenomena still await their experimental realization. In this
context, it has been theoretically predicted that, for homogeneous
systems, a multitude of phases appear on varying the interaction
parameters~\cite{marchetti}. In trapped systems, this gives rise
to states with a large variety of spatial
structures~\cite{molmer}.

As is known from experiments with single-component Bose and Fermi
systems, much physical information can be extracted by probing
collective motions. Theoretically, the collective excitations of
BF mixtures have been intensively studied for the phase-mixed BF
system~\cite{maruyama}, including the regime close to
phase-demixing~\cite{capuzzi}. In Ref.~\cite{sogo}, the collective
excitations of phase-segregated Bose-Fermi mixtures across phase
demixing were considered in case of very small fermionic particle
number.

In this work, we focus on the collective motions of isotropically
trapped, fully-segregated $^{40}$K-$^{87}$Rb mixtures. More
specifically, we investigate the eigenfrequencies as a function of
the fraction of the total particles that is fermionic, a parameter
which is accessible in experiments. Using an exact analytic
solution for the Boltzmann-Vlasov equation of a trapped,
collisionless Fermi gas, we investigate monopole modes and their
(collisionless) damping. To qualitatively investigate lower
symmetry (multipole) modes we use a fully hydrodynamic
description. This leads to the prediction of in-phase and
out-of-phase modes, with similar oscillations of inner and outer
boundaries. Whereas the breathing modes frequencies smoothly cross
over from the purely fermionic to the purely BEC spectrum, we find
that the presence of a small fraction of fermions has a dominant
effect on all other modes.

We consider in the following $N_B$ bosons and $N_F$ fermions,
trapped by isotropic harmonic confinement
$U_i(\mathbf{r})=m_i\omega_i^2r^2/2$ with $i=B,\,F$; we shall take
$m_B\omega_{B}^2=m_F\omega_F^2$ such that both particle species
feel the same external potential, as is the case for the
experiments on $^{40}$K and $^{87}$Rb
mixtures~\cite{ferlaino2,ospelkaus2,ferlaino3,zaccanti}. Within
mean-field theory the equilibrium state of a BF mixture with
chemical potentials $\mu_B$ and $\mu_F$, is fully characterized by
the wave function $\psi_B$ of the Bose-Einstein condensate (BEC)
and by the fermionic density $n_F$. The associated grand potential
is~\cite{molmer,marchetti}:
\begin{align*}
\Omega=&\int\text{d}\mathbf{r}\,
\psi_B^{*}\left(-\frac{\hslash^{2}\boldsymbol{\nabla}^2}{2m_B}
+U_B-\mu_B\right)\psi_B+\frac{G_{BB}}{2}|\psi_B|^4\\
&+(U_F-\mu_F)n_F+3G_{FF}n_F^{5/3}/5+G_{BF}n_F|\psi_B|^{2},\nonumber
\end{align*} with $G_{FF}=\hslash^{2}(6\pi^{2})^{2/3}/2m_F$, $G_{BF}=2\pi\hslash^2a_{BF}(m_F^{-1}+m_B^{-1})$
and $G_{BB}=4\pi\hslash^2a_{BB}/m_B$ the interspecies and bosonic
intraspecies coupling constant respectively. Minimization of this
grand potential with respect to $n_F(\mathbf{r})$ and
$\psi_B(\mathbf{r})$ gives the Thomas-Fermi (TF) and
Gross-Pitaevskii (GP) equations. When, however, the particle
number is sufficiently large, the ground state configuration is
well described using a local-density approximation. Therefore,
each point $\mathbf{r}$ in the trap can be treated locally, using
the effective chemical potentials
$\mu_i(\mathbf{r})=\mu_i-U_i(\mathbf{r})$, where $\mu_i\equiv
m\omega_{i}^2R_i^2/2$ the chemical potential at the center of the
trap and $i=B,\, F$.

Full BF phase segregation appears for sufficiently large
interspecies repulsion; more specifically, it occurs when $G_{BF}$
is larger than $G_{BB}\mu_F/\mu_B$ at the BF overlap region. For
all the $^{40}$K-$^{87}$Rb trap configurations considered here,
this requires the interspecies scattering length $a_{BF}$ to be
larger than $10^3$ Bohr, which is well below the experimentally
reported values of Ref.~\cite{ospelkaus}. The phase-segregated BF
trap then consists of a purely bosonic core surrounded by a purely
fermionic shell; in between there is a spherical shell where the
bosons and fermions overlap. The thickness of this shell is
generally very thin, of the order of the BEC healing length
$\hslash/(2m_B\mu_B)^{1/2}$. For almost all our further considered
configurations this length is much smaller than the trapping
length scales $\hbar/m_i\omega_i$ with $i=B,F$.

\begin{figure}
\begin{center}
  \epsfig{figure=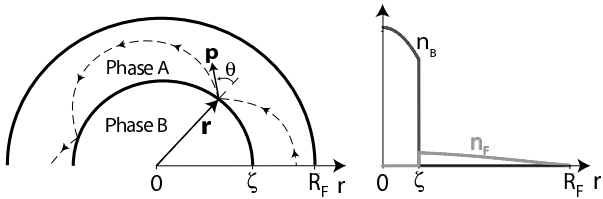,angle=0, height=65pt}
  \caption{(Left) Cross section of trap. Fermions and bosons are separated by the
    interface at radial position $\zeta$. Fermions follow
  trajectories along the dashed line and reflect off the
  interface. (Right) Boson and fermion densities $n_B$ and $n_F$ against radial
    position.
  \label{diagram}}
  \end{center}
\end{figure}

The influence of a thin interface on the equilibrium configuration
can be incorporated through an interface tension $\sigma$, an
exact expression for which is given in Eq.~\eqref{surface} in the
appendix. Mechanical equilibrium at the interface implies that the
pressures $P_i$ on either side of the interface are related by
Laplace's formula:
\begin{equation}\label{achilleasisadonkey}
P_B- P_F=2\sigma/\zeta,
\end{equation}
Here, $\zeta$ is the radial position of the interface (see
Fig.~\ref{diagram}). Finally the particle numbers of the BEC core
and fermion shell are:
\begin{subequations}\label{numbers}
\begin{align}
N_B&=\frac{R_B^5[m_B\omega_{B}/\hslash]^{2}}{2a_B}\left[\frac{\zeta^3}{3R_B^3}-\frac{\zeta^5}{5R_B^5}\right],\\
N_F&=\frac{[m_F\omega_{F}R_F^2/\hslash]^{3}}{72\pi
}[\mathcal{K}(1)-\mathcal{K}(\zeta/R_F)],
\end{align}
\end{subequations}
where $\mathcal{K}(x)\equiv
x\sqrt{1-x^2}(14x^2-3-8x^4)+3\arcsin(x)$.

The GP and TF equations, together with the number
equations~\eqref{numbers} and the condition for mechanical
equilibrium at the interface~\eqref{achilleasisadonkey} allow us
then to fully determine the equilibrium trap configuration.

Having established the equilibrium configuration of the BF system,
we now proceed to the description of its dynamics. We start by
discussing the dynamics of a BEC, which, as is well known, is
governed by the equations of hydrodynamics. Then, both
collisionless and hydrodynamical behavior is considered for the
outer Fermi shell. While a collisionless treatment is
experimentally more relevant, the succeeding hydrodynamic
consideration allows to bring out some important qualitative
aspects of the system's behavior. By combining the fermionic and
bosonic dynamics by means of the appropriate boundary conditions,
we then extract collective mode frequencies.

\section{Dynamics of the BEC}
As is usual, collective excitations are treated in a
hydrodynamical description by considering small-amplitude density
oscillations $\delta n_B$ which linearize the continuity,
$\partial_t\delta n_B+\boldsymbol{\nabla}\cdot(
n_B\mathbf{v}_B)=0$, and Euler, $m_B
n_B\partial_t\mathbf{v}_B=\boldsymbol{\nabla}\delta P_B- n_B
\boldsymbol{\nabla}U_B$, equations. Here, $n_B=\vert\psi_B\vert^2$
is the BEC condensate density. Using standard
methods~\cite{pethick,lazarides}, the chemical potential
oscillation $\delta\mu_B$ with temporal frequency $\omega$ then
satisfies:
\begin{equation}\label{genequation}
  m_B n_B\omega^2\delta\mu_B+\mu_B\boldsymbol{\nabla}\cdot\left(
  n_B\boldsymbol{\nabla}\delta\mu_B\right)=0,
\end{equation}
the solution to which is $\delta\mu_B\propto
\mathrm{Y}_{\ell}^mr^\ell
F(\alpha^+,\alpha^-,\ell+3/2,(r/\tfrad{B})^2)$, with $\text{F}$
the hypergeometric function, $\mathrm{Y}_{\ell}^m$ the spherical
harmonic,
$2\alpha^{\pm}=z\pm[z^2+2(\omega^2/\omega^2_B-\ell)]^{1/2}$,
$z=\ell+1/2+1$ and $\ell$ a positive integer or zero.

Here, we must impose a boundary condition at the BF interface.
Since we are assuming that the oscillations occur at local
thermodynamic equilibrium~\footnote{This is why we can use the
equation of state, for example.}, we conclude that the two fluids
are completely immiscible and the interface impermeable. This
immediately leads to the condition
\begin{equation}\label{impermeability}
    \dot{\zeta}=\boldsymbol{e}_\zeta\cdot\left.\mathbf{v}_B\right|_{\zeta},
\end{equation}
in which $\boldsymbol{e}_\zeta$ is a unit radial vector normal to
the interface, and the dot denotes the derivative with respect to
time. That is, the interface velocity must be equal to the BEC
velocity at the interface, ensuring vanishing of the flux of
bosons through the interface.

\section{Collisionless Fermi Gas}
If the fermionic particles
are collisionless, as expected at low temperatures, the
appropriate dynamical description is the Boltzmann-Vlasov equation
with a vanishing collision integral:
\begin{equation}\label{eq:boltzmann}
\partial_t f+\boldsymbol{v}\cdot\boldsymbol{\nabla}_\mathbf{r}
f-\omega_F^2\mathbf{r}\cdot\boldsymbol{\nabla}_\mathbf{v} f=0.
\end{equation}
The mean-field interaction term is also absent, again because
fermionic interactions are suppressed by Pauli blocking.  Since
collisions do not act to restore local equilibrium during the
oscillation, the Fermi sphere is deformed during the oscillation.
We therefore write
$f=f_0+\delta(|\boldsymbol{p}|-\pf)\cdot\nu(r,\chi)\mathrm{e}^{-i\omega
  t}$, where $\pf(\mathbf{r})=[2m_F\mu_F(\mathbf{r})]^{1/2}$ is the local Fermi momentum and
$\chi=\cos\theta$ with $\theta$ the angle between $\mathbf{r}$ and
$\mathbf{p}$ (see Fig.1). This parametrization is the most
convenient for studying spherically symmetric oscillations.
Eq.~\eqref{eq:boltzmann} then becomes
\begin{equation}\label{eq:vlasov}
-i\omega\nu+\omega_F \widetilde{p}\chi\,\partial_r\nu
+\omega_F(1-\chi^2)g(r)\partial_\chi\nu=0,
\end{equation}
with $g(r)=\widetilde{p}/r-r/\widetilde{p}$ and
$\widetilde{p}^2\equiv\tfrad{F}^2-r^2$. The general solution to
Eq.~\eqref{eq:vlasov} can readily be found:
\begin{equation}\label{eq:nugeneral}
\nu(r,\chi)=\mathcal{C}[r^2(\tfrad{F}^2-r^2)(1-\chi^2)]e^{i\omega\tau/2},
\end{equation}
with $\mathcal{C}[x]$ an arbitrary function of the variable
$x$~\footnote{Note that the paths along which the squared angular
momentum $r^2(R_F^2-r^2)(1-\chi^2)m_F^2\omega_F^2$ is constant,
corresponds to classical orbits in a harmonic potential.}, and
\begin{equation}\label{eq:tau}
\tau(r,\chi)=\arctan[2\chi/g(r)]/\omega_F.
\end{equation}
Physically, $\tau(\zeta,\chi)$ is the time for a particle, leaving
the interface at an angle $\theta=\arccos(\chi)$ and speed
$p_f(\zeta)/m_F$, to return to it.

To fix the arbitrary function $\mathcal{C}$ in the general
solution of Eq.~\eqref{eq:nugeneral}, we must once more impose the
appropriate boundary condition. As already explained, the
interface is completely impermeable to both particle types;
incident fermions are therefore specularly
reflected~\cite{vanschaeybroeck}. A boundary condition encoding
this process may be derived using the following arguments, after
Bekarevich and Khalatnikov~\cite{bekarevich}: first, in a frame
moving with the interface, the energy of the incident and
reflected excitations must be equal; second, the influx of
particles in this frame must be equal to their outflux. These two
conditions result in~\cite{lazarides}:
\begin{equation}\label{eq:bkbc}
\nu(\zeta,\chi)-\nu(\zeta,-\chi)=2m_F \chi\dot{\zeta},
\end{equation}
where $\dot\zeta$ is the velocity of the interface. One may
understand Eq.~\eqref{eq:bkbc} as follows: for a stationary
interface, $\dot{\zeta}=0$, all particles incident at angle
$\theta$ will be reflected such that $\nu(\chi)=\nu(-\chi)$. If,
on the other hand, the interface is moving with $\dot{\zeta}>0$,
then $\nu(\chi)-\nu(-\chi)$ is positive and proportional to
$\chi$; that is, a) the reflected state at any given angle becomes
more highly occupied than the incident state at the same angle,
and b) the difference in occupancy is largest when $\chi=1$, that
is, for particles incident normal to the interface. Thus, there is
an overall probability transfer to states closer to normal
($\chi=1$). This is due to the momentum transfer from the
interface to the reflected particles. At $r=\zeta$, the solution
of the Boltzmann-Vlasov Eq.~\eqref{eq:nugeneral} which satisfies
condition~\eqref{eq:bkbc} is:
\begin{equation}\label{eq:nusoln}
  \nu\left(\zeta,\chi\right)=
  m_F\chi\dot{\zeta}\left[1-i\cot(\omega\tau/2)\right].
\end{equation}

Demanding local mechanical equilibrium at the interface results in
the linearized Laplace equation with the role of the fermionic
pressure fluctuation played by the radial component of the
momentum flux tensor
$\delta\Pi_{rr}^F=\int\mathrm{d}^3\mathbf{p}\,(\chi
p)^2(f-f_0)/[m_F(2\pi\hslash)^3]$. Thus,
\begin{align}\label{eq:laplacecollisionless}
&n_B\delta\mu_B-\delta\Pi_{rr}^F=
\sigma\delta\zeta(\ell-1)(\ell+2)/\zeta^2\\
&-3\sigma\delta\zeta
m\omega_F^2/\mu_B+3\sigma\delta\mu_B/(\mu_B\zeta)-\delta\zeta\partial_{r}\left(P_B-P_F\right),\nonumber
\end{align}
with all quantities evaluated at the interface and
$\delta\zeta\propto\text{Y}_{\ell}^{m}(\theta,\phi)$ the departure
of the interface from its equilibrium position. Solving
Eqs.~\eqref{eq:laplacecollisionless} and \eqref{impermeability}
for $\omega$ fixes the eigenfrequencies.

\begin{figure}
  \epsfig{figure=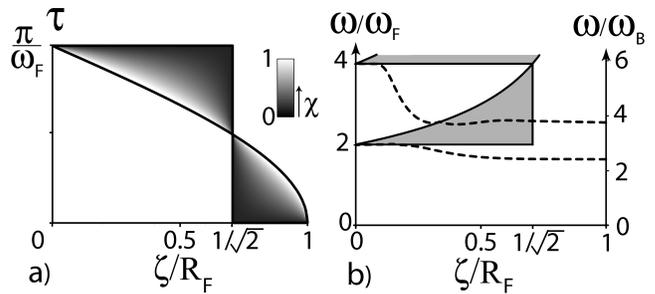,angle=0, width=240pt}
  \caption{(a) The full lines outline the region of possible values of $\tau$ (see
  Eq.\eqref{eq:tau}) as a function of the interface position
  $\zeta$. (b) For any multipole mode, damping will occur for values of $\Re(\omega)$ and $\zeta$ in the shaded region.
  The dashed lines correspond to the monopole modes depicted in Fig.~\ref{fig:collessfreqs} with full lines.
  \label{fig:damping}
  }
\end{figure}
\section{Damping} Despite the absence of collisions in the
fermionic gas, damping of the collective oscillation may occur,
induced by the interaction of the fermions with the interface. To
understand its origin, consider the two time scales relevant in
the collisionless fermionic phase: the period of oscillation
$2\pi/\omega$, and $\tau$, the time of flight for a particle
projected at angle $\theta=\arccos(\chi)$ from the interface (see
Eq.~\eqref{eq:tau}). From Eq.~\eqref{eq:nusoln}, $\nu(\zeta,\chi)$
diverges if:
\begin{align}\label{eq:dampingcondition}
\tau(\zeta,\chi)=2\pi n/\omega,
\end{align}
with $n$ an integer; that is, when there exist particles which
depart from and return to the interface in exactly an integer
number of oscillation periods. The resulting pole on the
integration path in the pressure tensor $\delta\Pi_{rr}^F$ and
thus also in the dispersion relation
Eq.~\eqref{eq:laplacecollisionless}, must be avoided by some
definite prescription~\cite{landau}. The appropriate manner is by
deforming the contour of integration for $\delta\Pi_{rr}^F$ so as
to pass below the pole~\footnote{The validity of this method can
be justified by repeating the calculations using Laplace, rather
than Fourier, transforms in time.}. This gives rise to an
additional, imaginary term due to the residue in
$\delta\Pi_{rr}^F$; in this way, the solution for $\omega$ to the
dispersion relation acquires a negative imaginary part: damping
occurs.

Can one intuitively understand when damping occurs? In other
words, under which conditions are there particles which satisfy
Eq.~\eqref{eq:dampingcondition}? As we will see, whether damping
occurs or not depends solely on the interface position and the
real part of $\omega$, $\Re(\omega)$, and this remains true for
the multipole modes.

For fixed interface position $\zeta/\tfrad{F}$, the time $\tau$
takes values only within a restricted
interval~(Fig.~\ref{fig:damping}a). Consider first radially moving
particles ($\chi=1$); for vanishing $\zeta/\tfrad{F}$, they take
half a trapping period to return to the interface, i.e.,
$\tau(0,1)=\pi/\omega_F$ and, as $\zeta$ is increased, $\tau$
smoothly vanishes as the interface approaches the trap edge.
For $\zeta/\tfrad{F}>1/\sqrt{2}$, particles projected parallel to
the interface ($\chi=0$) experience a ``centrifugal force'' which
is insufficient to overcome the harmonic trapping force such that
$\tau(\zeta,0)=0$; for general $\chi$, one finds
$0<\tau(\zeta,\chi)<\tau(\zeta,1)$. Conversely, if
$\zeta/\tfrad{F}<1/\sqrt{2}$, a particle with $\chi=0$ performs
exactly half an orbit between two collisions with the interface
and therefore $\tau(\zeta,0)=\pi/\omega_F$; in this case,
$\tau(\zeta,1)<\tau(\zeta,\chi)<\pi/\omega_F$ for general $\chi$.

From the possible values of $\tau$, it is straightforward to
obtain the values of $\omega$ and $\zeta$ for which damping occurs
(see shaded region in Fig.~\ref{fig:damping}b). For small
$\zeta/\tfrad{F}$, we have seen that $\tau\approx \pi/\omega_F$;
but this is exactly half the period of a collisionless Fermi gas.
Damping is therefore expected for the lowest monopole mode when
$N_B\ll N_F$. On the other hand, when $\zeta/\tfrad{F}\approx 1$,
due to a vanishing $\tau$, Eq.~\eqref{eq:dampingcondition} is
satisfied only for large values of $\omega$. This inhibits damping
for the lowest modes in case $N_F\ll N_B$.

The damping we observe is somewhat analogous to Landau damping,
which corresponds to energy transfer from collective modes into
single-particle incoherent modes.  In our case, no clear
distinction exists between collective and single-particle modes in
the Fermi gas and energy goes into incoherent excitations in the
collisionless gas.

\begin{figure}
  \epsfig{figure=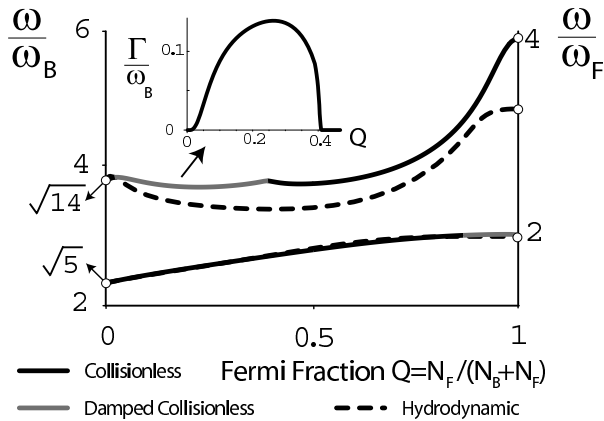,angle=0, width=200pt}
  \caption{The frequencies $\omega$ of the $\ell=0$ monopole modes as a function of the Fermi fraction
$Q=N_F/(N_B+N_F)$. The full black and grey lines correspond to
damped and undamped monopole modes, respectively, both calculated
using collisionless fermion dynamics. The inset shows the damping
rate $\Gamma$ of the second monopole mode. Dashed lines are
calculated using a hydrodynamic Fermi shell.
\label{fig:collessfreqs}
  }
\end{figure}

\section{Results} We consider trap configurations containing $10^6$
particles and characterized by the parameters $a_{BB} = 5.23$ nm
and $\omega_B=570$ Hz~\footnote{This frequency is the geometrical
mean of the experimental frequencies of
Ref.~\onlinecite{ospelkaus2}.}. In Fig.~\ref{fig:collessfreqs}, we
show with full lines the real part of the frequencies of the two
lowest monopole modes as a function of the Fermi fraction
$Q=N_F/(N_B+N_F)$. The grey and black lines indicate the damped
and undamped frequencies respectively. The lowest monopole mode
smoothly crosses over between the value $\sqrt{5}\omega_B$ for a
fully superfluid bosonic trap at $Q=0$, to the value $2\omega_F$
of the fully collisionless (fermionic) trap at
$Q=1$~\cite{pethick}. As explained earlier, the lowest monopole
mode is damped when $\zeta/R_F\approx 0$; indeed, for $Q>0.85$,
very weak damping is present with a maximum damping rate of about
$10^{-3}\omega_B$. As expected, the second monopole mode frequency
goes over from $\sqrt{14}\omega_B$ at $Q=0$ to $4\omega_F$ at
$Q=1$~\cite{pethick}. Its variation, however, is nonmonotonic and
considerable damping occurs in the interval $0.02<Q<0.4$; the
associated damping rate $\Gamma$ is shown in the inset of
Fig.~\ref{fig:collessfreqs}. The absence of damping for low values
of $Q$ is in full agreement with the explanation given above. In
Fig.~\ref{fig:damping}b, we plot the evolution of $\omega$ as a
function of $\zeta$ (dashed lines) for both monopole modes;
damping occurs in the shaded region.

\section{Out-of-Phase Modes}
 While a collisionless treatment is
experimentally more relevant, the following hydrodynamic
considerations for the fermions allow us to bring out some
important qualitative aspects of the system's dynamical behavior.

The bulk solutions to the continuity and Euler equations for
fermions are similar to those for the BEC~\cite{lazarides}. The
Laplace Eq.~\eqref{eq:laplacecollisionless} used for the
collisionless case still holds but with $\delta\Pi_{rr}^F$
replaced by $n_F \delta\mu_F$ and the impermeability of the
interface implies
$\dot{\delta\zeta}=\boldsymbol{e}_\zeta\cdot\left.\mathbf{v}_F\right|_{\zeta}$.

\begin{figure}
  \epsfig{figure=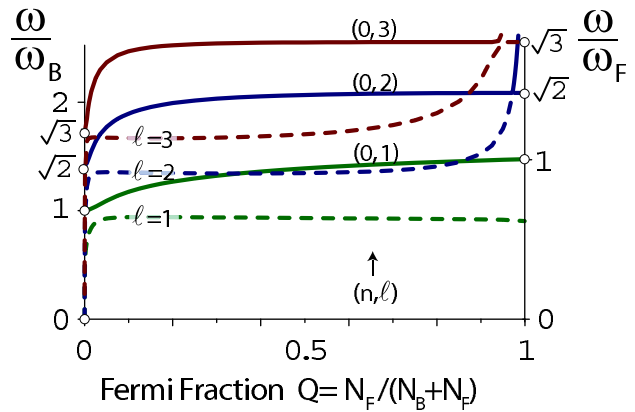,angle=0, width=220pt}
  \caption{The frequencies $\omega$ of the multipole ($\ell\neq 0$) modes as a function of the Fermi fraction
$Q=N_F/(N_B+N_F)$. All are found using a fully hydrodynamical trap
and involve in-phase (full lines) and out-of-phase (dashed lines)
motions of inner and outer boundary. \label{fig:hydrofreqs}
  }
\end{figure}

In Fig.~\ref{fig:collessfreqs}, we show with dashed lines the
frequencies for the breathing modes $\ell=0$ using a
hydrodynamical Fermi shell. In Fig.~\ref{fig:hydrofreqs}, on the
other hand, the lowest multipole ($\ell\neq 0$) modes are shown.
Both the multipole modes in which the interface and outer boundary
move in phase (full lines) and out of phase (dashed lines), are
present~\cite{lazarides}. Both the IP and OOP modes have in-phase
velocities at the BF interface and are both generic to a
two-component system~\cite{svidzinsky,lazarides}. An interesting
result is that the frequencies of all the $\ell\neq 0$ modes vary
strongly at small $Q$ and remain approximately constant at larger
$Q$; this in contrast with the monopole modes.

\section{Discussion}
The mass difference between the $^{40}$K
and $^{87}$Rb particles may cause a gravitational sag to be
present. Incorporating such effect is beyond the scope of this
work. However, we expect gravity not to affect the validity of our
qualitative results such as the damping mechanism and the presence
of out-of-phase modes. Note also that the gravitational sag may be
quenched experimentally by counterbalancing external potentials or
by tight radial confinement.

A finite temperature will change the collective mode frequencies
mostly through a change in the fermion dynamics. We expect
$\Re(\omega)$ to be changed little with temperature, while damping
is expected to occur for a broader range of $Q$ and occurs by the
same mechanism as described here~\cite{vanschaeybroeck2}

Our approach fails when the interface is as thin as the outer
fermion shell; we find that this happens for Fermi fractions below
$2\times 10^{-3}$. In that case, damping of the collective
excitations happens by usual Landau damping throughout the BF
overlap region~\cite{sogo}. Our work is complementary to
Ref.~\cite{sogo} in the sense that they study the process of phase
demixing for a Fermi fraction $4\times 10^{-5}$ whereas we focus
on the variation of the Fermi fraction in case of complete phase
segregation.

\section{Conclusion}
We present monopole mode frequencies for
the trapped phase-segregated $^{40}$K-$^{87}$Rb mixture. A damping
mechanism is identified which is generic to all collective modes.
Within a fully hydrodynamic approximation we then obtain mode
frequencies for modes involving out-of-phase and in-phase motions
of inner and outer boundaries. All our results are only very
weakly dependent on the surface tension.

\section{Acknowledgement} We acknowledge partial support by Project
No.~FWO G.0115.06; A.L.~is supported by Project No.~GOA/2004/02
and B.V.S. by the Research Fund K.U.Leuven.

\section{Appendix}
In order to obtain the surface tension of a BF interface, one may
consider an infinite space containing fermions at
$x\rightarrow-\infty$ and bosons at $x\rightarrow+\infty$, bound
by a flat interface which is parallel to the $y-z$ plane. Both
species are taken to be untrapped ($U_B=U_F=0$ and at fixed
chemical potentials $\mu_B$ and $\mu_F$. The appropriate boundary
conditions for $\psi_B(x)$ and $n_F(x)$ are
$\psi_B(-\infty)=n_F(+\infty)=0$ while
$\psi_B(+\infty)=\sqrt{\mu_B/G_{BB}}$ and
$n_F(-\infty)=(\mu_F/G_{FF})^{3/2}$. We define the BEC coherence
length $\xi_B\equiv\hslash/[2m_B\mu_B]^{1/2}$ and the
dimensionless parameters $\kappa^2\equiv G_{BB}\mu_F/G_{BF}\mu_B$
and $\phi_B(x)\equiv \psi_B(x)/\sqrt{\mu_B/G_{BB}}$. Using these
variables, is useful to introduce the first integral associated
with the GP and TF equations for our system under consideration:
\begin{align}\label{surface1}
2\xi_B^2(\partial \phi_B/\partial
x)^2=(1-\phi_B^2)^2-\mathcal{G}(\kappa,\phi_B),
\end{align}
where
$\mathcal{G}(\kappa,\phi_B)=(1-\phi_B^2/\kappa^2)^{5/2}\Theta(\kappa-\phi_B)$
and $\Theta$ the Heaviside function. Using Eq.~\eqref{surface1},
the interface tension which is the excess grand potential per unit
area of the interface can be recast in the form:
\begin{align*}
\sigma=&4P\xi_B^2\int_{-\infty}^{\infty}\text{d}x\,
(\partial\phi_B/\partial x)^2,
\end{align*}
Using the GP and TF equations and the transformation
$\text{d}x=(\partial \phi_B/\partial x)^{-1}\text{d}\phi_B$, the
interface tension $\sigma$ associated with a flat BF interface is:
\begin{align}\label{surface}
\sigma(\kappa)&=2\sqrt{2}P\xi_B\int_{0}^{1}\text{d}
\phi_B\sqrt{(1-\phi_B^2)^2-\mathcal{G}(\kappa,\phi_B)},
\end{align}
The surface tension is minimal at $\kappa=1$ ($\sigma=0.52P\xi_B$)
and reaches its maximal value $4\sqrt{2}P\xi_B/3$ when $\kappa=0$.
Note that one may use this exact result for the surface tension in
a trap when the chemical potentials $\mu_B$ and $\mu_F$ are
evaluated at the interface.

\end{document}